\documentclass{article}
\usepackage[T1]{fontenc} 
\usepackage[utf8]{inputenc} 
\usepackage{ismir,amsmath,cite,url}
\usepackage{graphicx}
\usepackage{color}
\usepackage{lineno}

\usepackage{bm}
\usepackage{amssymb}
\newcommand{\V}[1]{{\bf #1}} 
\newcommand{\Vg}[1]{\bm{#1}} 
\def \RR {\mathbb{R}} 

\title{A Kalman Filter model for synchronization in musical ensembles}

\multauthor
{Hugo T. Carvalho$^{1*}$ \hspace{1cm} Min S. Li$^2$ \hspace{1cm} Massimiliano di Luca$^3$ \hspace{1cm} \bfseries{Alan M. Wing$^3$}} {$^1$ Department of Statistical Methods, Federal University of Rio de Janeiro, Brazil\\
$^2$ Bristol Interaction Group, School of Computer Science, University of Bristol, United Kingdom \\
$^3$ Virtual Reality Lab, School of Psychology, University of Birmingham, United Kingdom\\
{\small $^*$ Corresponding author: \tt hugo@dme.ufrj.br}
}

\def\authorname{H. T. Carvalho, M. S. Li, M. di Luca, and A. M. Wing}

\begin{document}

\maketitle
\begin{abstract}
The synchronization of motor responses to rhythmic auditory cues is a fundamental biological phenomenon observed across various species. While the importance of temporal alignment varies across different contexts, achieving precise temporal synchronization is a prominent goal in musical performances. Musicians often incorporate expressive timing variations, which require precise control over timing and synchronization, particularly in ensemble performance. This is crucial because both deliberate expressive nuances and accidental timing deviations can affect the overall timing of a performance. This discussion prompts the question of how musicians adjust their temporal dynamics to achieve synchronization within an ensemble. This paper introduces a novel feedback correction model based on the Kalman Filter, aimed at improving the understanding of interpersonal timing in ensemble music performances. The proposed model performs similarly to other linear correction models in the literature, with the advantage of low computational cost and good performance even in scenarios where the underlying tempo varies.
\end{abstract}
\section{Introduction}\label{sec:introduction}
Synchronization of motor responses to rhythmic auditory cues represents a biological phenomenon found across various species \cite{patel-etal}, and social collectives often engage in activities necessitating precise temporal coordination among members, a crucial factor for successful group endeavors. For example, in scenarios such as rowing eights, temporal alignment may not be the primary focus, but individual timing remains tied to collective timing dynamics \cite{wing-woodburn}. In domains like musical performances, achieving precise temporal synchronization serves as a prominent goal \cite{goodman}. 

Typically, musicians do not adhere strictly to the exact timing of note onsets as indicated in the musical score: due to expressiveness, they often introduce deviations from the prescribed timing \cite{palmer}. These fluctuations require a high level of control over relative timing, where the phase of notes produced by the musician deviating from the timing aligns differently with the phases of other musicians. Rehearsals often involve reaching a consensus on expressive variations, ensuring that timing deviations are synchronized among players while maintaining relative timing \cite{davidson-good}. Nevertheless, even with a unified understanding of the musical interpretation, individual musicians may opt to vary the timing of note onsets in specific passages between different performances \cite{davidson-good, murninghan-conlon}. Musical performance timing is also susceptible to inadvertent variations due to factors such as rhythmic intricacies, technical demands beyond timing (e.g., pitch, volume), lapses in concentration, and the inherent variability of biological timing \cite{gibbon-et-al}. While extensive individual practice can mitigate some of these unintended variations, complete elimination is unlikely. 

The previous discussion raises the inquiry: how do musicians within an ensemble modulate their temporal dynamics to achieve synchronization with one another? In this paper a novel feedback correction model is presented, based on the Kalman Filter and aimed at improving timing accuracy in ensemble music performances. The proposed model generalizes the linear autoregressive model in \cite{wing-etal-2014} with the improvement of allowing two important quantities, the \textit{phase} and \textit{period} \textit{correction gains}, to vary along time, since it makes the model suitable to describe synchronization in scenarios where the underlying tempo greatly varies (a realistic case in ensemble performance). 

The paper is organized as follows: Section \ref{sec:lin-models} recalls some linear models for synchronization, and the dynamic generalization of the model in \cite{wing-etal-2014} is presented, which is formulated as a Kalman Filter in Section \ref{sec:kf-model}; the fundamentals of the Kalman Filter are briefly recalled in Section \ref{sec:kf-recall}; the computational experiments are shown and discussed in Section \ref{sec:results}; conclusions are presented in Section \ref{sec:conc-fut-works}. Directions for future work are identified throughout the paper.
\section{Linear models for ensemble synchronization}\label{sec:lin-models}
The starting point for contextualizing the proposed model is \cite{vorberg-schulze}, where a phase-correction model is presented as a method for an individual
performer to achieve synchrony with a periodic metronome click or with another performer (see also \cite{vorberg-wing}). The fundamental concept revolves around utilizing the asynchrony, termed as a \textit{phase error}, between a tone onset and the metronome click (or between two tone onsets produced by different performers) in a feedback mechanism, that guides the performer to adjust the time interval preceding the next tone onset. Consequently, the performer either decreases or increases the interval leading up to the subsequent tone onset proportionally to the preceding asynchrony. This process aims to achieve greater synchrony (``in phase'') between the next tone onset and the metronome click (or pair of tone onsets). This synchronization scheme can be represented by Equation \eqref{eq:vorberg-schulze-model}:
\begin{align}
    t_n = t_{n - 1} + T_n - \alpha A_{n - 1} + \varepsilon_n,
    \label{eq:vorberg-schulze-model}
\end{align}
where $t_n$ and $t_{n-1}$ represent the current and previous observed tone onset event times respectively, $T_n$ denotes the time interval generated by an internal timekeeping mechanism, $\alpha$ is the \textit{phase correction strength} or \textit{phase correction gain}, $A_{n-1}$ refers to the asynchrony of the previous onset event, and $\varepsilon_n$ represents a random error term, which includes the internal timekeeper noise. 
The complete reduction of asynchrony to zero hinges on the value assigned to the gain, $\alpha$, since this parameter determines the proportion of the preceding phase error that the performer endeavors to eliminate.

Following \cite{mates}, \textit{period correction} can also be incorporated in the model in Equation \eqref{eq:vorberg-schulze-model}, by imposing that
\begin{align}
    T_n = T_{n - 1} - \beta A_{n - 1},
    \label{eq:period-correction}
\end{align}
where $\beta$ is the \textit{period correction strength} or \textit{period correction gain}. Phase correction involves a local, within-cycle adjustment to the timing, while period correction entails a more enduring alteration to the underlying tempo, influencing subsequent cycles as well. Phase correction typically occurs automatically, without the need for conscious awareness of synchronization discrepancies. However, period correction appears to be more cognitively demanding, relying on the conscious detection of tempo variations in the external rhythm \cite{repp-adap-tempo-changes, repp-keller-adap-tempo-changes}.

As previously mentioned, Equations \eqref{eq:vorberg-schulze-model} and \eqref{eq:period-correction} model the asynchrony correction of an individual tapping according to a periodic metronome, or between two individuals tapping together. In \cite{wing-etal-2014} it is argued that the same modeling framework is also suited to describe synchronization in music ensemble performance, where a specific musician now tries to reduce asynchrony between him/her and every other performer. Therefore, Equations \eqref{eq:vorberg-schulze-model} and \eqref{eq:period-correction} can be jointly generalized to an ensemble of $K$ performers as:
\begin{align}
    t_{i, n} &= t_{i, n - 1} + T_{i, n} - \sum_{\substack{i = 1 \\ j \neq i}}^{K} \alpha_{ij}A_{ij, n - 1} + \varepsilon_{i, n} \label{eq:wing-2014-model-t} \\
    T_{i, n} &= T_{i, n - 1} - \sum_{\substack{i = 1 \\ j \neq i}}^{K} \beta_{ij}A_{ij, n - 1}, \label{eq:wing-2014-model-T} 
\end{align}
where $i = 1, \dots, K$ indicate a specific performer, $t_{i, n}$ and $t_{i, n - 1}$ are respectively the current and previous observed tone onset event times for player $i$, $T_{i, n}$ is the timekeeper interval for player $i$ at time instant $n$, $A_{ij, n - 1} = \left(t_{i, n - 1} - t_{j, n - 1} \right)$ is the asynchrony at the time instant $n - 1$ between players $i$ and $j$, $\alpha_{ij}$ and $\beta_{ij}$ are respectively the phase and period correction gain applied by player $i$ to compensate for $A_{ij, n - 1}$, and $\varepsilon_{i, n}$ is a noise term identified with the internal timekeeper. Estimation of the values of $\alpha_{ij}$ and $\beta_{ij}$ can be performed using the \textit{bounded Generalized Least Squares} method (bGLS) \cite{jacoby-bgls-1, jacoby-bgls-2}. 

In \cite{wing-etal-2014}, the model in Equation \eqref{eq:wing-2014-model-t} is implemented and largely investigated for the case of a string quartet ensemble playing a homophonic section from the string quartet Op. 74 no. 1 by Joseph Haydn (fourth movement, bars 13--24), as this part has a steady tempo and all player's quarter notes are aligned. In \cite{jacoby-bgls-1} the coupling of Equations \eqref{eq:wing-2014-model-t} and \eqref{eq:wing-2014-model-T} is investigated, with a simulated string quartet data with mild tempo changes, and the bGLS algorithm is shown to be capable of recovering the values of $\alpha$ and $\beta$. However, due to the nature of the bGLS algorithm, the authors point out that many data points are necessary for robust estimation of these variables, which may not be available or is an unrealistic aim in the case of a real-time implementation of the correction model (eg. for a virtual reality musical ensemble). In \cite{adam-2013} the ADAM model (ADaptation and Anticipation Model) is proposed, including not only correction terms but also anticipatory ones, and in \cite{adam-2019} this model is tested with tempo-changing tapping data, but since there is no adaptation of the bGLS algorithm to this new set of equations, the parameter estimation is done by exhaustive search, which is infeasible for real-life applications. Moreover, due to the nature of its parameters, the ADAM model is non-identifiable, meaning that more than one configuration of the parameters leads to the same estimate.

In order to circumvent the aforementioned issues, an alternative is to consider not a single value of $\alpha$ and $\beta$ for each pair of performers through time, but \textit{time-dependent correction gains}. Developing this intuition, a dynamic $\alpha_{ij}$ allows that a performer changes the phase correction at each onset, and a dynamic $\beta_{ij}$ would allow him/her to correct differently for tempo variations during the performance of an excerpt. To model a dynamic variable, a good balance between simplicity and accuracy is a random walk, and in this case phase and period correction occur according to Equations \eqref{eq:wing-2014-model-t} and \eqref{eq:wing-2014-model-T}, respectively, but with additional equations to allow the evolution of both correction gains. This new model is summarized in Equations \eqref{eq:wing-2014-model-dynamic-ab-t}, \eqref{eq:wing-2014-model-dynamic-ab-T}, \eqref{eq:wing-2014-model-dynamic-ab-alpha}, and \eqref{eq:wing-2014-model-dynamic-ab-beta}:
\begin{align}
    t_{i, n} &= t_{i, n - 1} + T_{i, n} - \sum_{\substack{i = 1 \\ j \neq i}}^{K} \alpha_{ij, n} A_{ij, n - 1} + \varepsilon_{i, n} \label{eq:wing-2014-model-dynamic-ab-t} \\
    T_{i, n} &= T_{i, n - 1} - \sum_{\substack{i = 1 \\ j \neq i}}^{K} \beta_{ij, n} A_{ij, n - 1} \label{eq:wing-2014-model-dynamic-ab-T}  \\
    \alpha_{ij, n} &= \alpha_{ij, n - 1} + w_{ij, n}^{(\alpha)} \label{eq:wing-2014-model-dynamic-ab-alpha} \\
    \beta_{ij, n} &= \beta_{ij, n - 1} + w_{ij, n}^{(\beta)}, \label{eq:wing-2014-model-dynamic-ab-beta}
\end{align}
where $w_{ij, n}^{(\alpha)}$ and $w_{ij, n}^{(\beta)}$ are independent zero-mean Gaussian random variables, allowing the evolution of $\alpha_{ij, n}$ and $\beta_{ij, n}$ through time, respectively (notice the novel subscript ``$n$''~in both $\alpha_{ij}$ and $\beta_{ij}$).

However, in the model proposed in Equations \eqref{eq:wing-2014-model-dynamic-ab-t}, \eqref{eq:wing-2014-model-dynamic-ab-T}, \eqref{eq:wing-2014-model-dynamic-ab-alpha}, and \eqref{eq:wing-2014-model-dynamic-ab-beta}, it is not clear how to employ the bGLS method to obtain estimate of the variables of interest, and two distinct paths can now be followed: generalize the bGLS algorithm to this new situation, or resort to estimation techniques within the theory of dynamic models \cite{petris-dyn-models-r}. This work follows the latter, adopting the Kalman Filter as a framework to analyze Equations \eqref{eq:wing-2014-model-dynamic-ab-t}, \eqref{eq:wing-2014-model-dynamic-ab-T}, \eqref{eq:wing-2014-model-dynamic-ab-alpha}, and \eqref{eq:wing-2014-model-dynamic-ab-beta}, due to its balance between flexibility and simplicity, as well as its simple and highly interpretable update equations. Section \ref{sec:kf-recall} recalls the basics of the Kalman Filter and Section \ref{sec:kf-model} formulates the proposed model in this scenario.

\section{A brief recall on the Kalman Filter}\label{sec:kf-recall}
The Kalman Filter (KF) was developed in the 1960's, and served originally as a way to produce accurate estimates of variables of interest (eg. position of an object) by reaching a consensus between physical models and noisy measurements \cite{grewal-andrews-kf}. More generally, the KF can be seen as a state-space dynamic model, employed to describe more general time-series as a dynamic linear regression model as function of an underlying Markov model \cite{petris-dyn-models-r}.  

The main contribution of this paper is to propose the model in Equations \eqref{eq:wing-2014-model-dynamic-ab-t}, \eqref{eq:wing-2014-model-dynamic-ab-T}, \eqref{eq:wing-2014-model-dynamic-ab-alpha}, and \eqref{eq:wing-2014-model-dynamic-ab-beta}, and formulate it as a KF, employing its filtering and smoothing equations to estimate the phase and period correction gains through time. The choice of a KF to achieve this goal are: linear nature of the model in Equations \eqref{eq:wing-2014-model-dynamic-ab-t}, \eqref{eq:wing-2014-model-dynamic-ab-T}, \eqref{eq:wing-2014-model-dynamic-ab-alpha}, and \eqref{eq:wing-2014-model-dynamic-ab-beta}, high interpretability of the KF and its update equations, and potential low computational cost of its implementation.

The notation and basic equations of the KF are now briefly recalled, following \cite{petris-dyn-models-r}. In what follows, the index $n$ ranges from $1$ to $N$. Let $\V{y}_n \in \RR^m$ be a sequence of \textit{observed variables} (or \textit{measurements}), and $\Vg{\theta}_n \in \RR^p$ be a sequence of unobserved vectors, which are called the \textit{hidden} (or \textit{state}) variables. The KF model assumes that these two entities are related by Equations \eqref{eq:kf-obs} and \eqref{eq:kf-hidden}:
\begin{align}
    \V{y}_n &= \V{F}_n\Vg{\theta}_n + \V{v}_n \label{eq:kf-obs} \\
    \Vg{\theta}_n &= \V{G}_n\Vg{\theta}_{n - 1} + \V{w}_n, \label{eq:kf-hidden}
\end{align}
where $\V{F}_n \in \RR^{m \times p}$ and $\V{G}_n \in \RR^{p \times p}$ are sequences of known matrices (\textit{observation model} and the \textit{state-transition model}, respectively). Vectors $\V{v}_n \in \RR^m$ and $\V{w}_n \in \RR^p$ are independent \textit{observation} and \textit{process} noise terms, respectively, and it is assumed that they follow Gaussian probability distributions, that is, $\V{v}_n \sim N(\V{0}, \V{V}_n)$ and $\V{w}_n \sim N(\V{0}, \V{W}_n)$,\footnote{The symbol $\sim$ means ``follows the probability distribution'', and $N(\Vg{\mu}, \Vg{\Sigma})$ denotes a multivariate Gaussian distribution with mean vector $\Vg{\mu}$ and covariance matrix $\Vg{\Sigma}$. The dimension of the support of the random vector is omitted, and compatibility between dimensions of $\Vg{\mu}$ and $\Vg{\Sigma}$ is always assumed.} where $\V{V}_n \in \RR^{m \times m}$ and $\V{W}_n \in \RR^{p \times p}$ are sequences of known covariance matrices of the observation and process noise terms respectively.

The KF dynamically estimates variables $\Vg{\theta}_n$ and $\V{y}_n$ based on observations up to time $n - 1$, and updates the estimate of $\Vg{\theta}_n$ when the observation at time $n$ is available. This process is done accordingly to Equations \eqref{eq:kf-pred-hid}, \eqref{eq:kf-pred-obs} and \eqref{eq:kf-compare-w-meas}, called the \textit{filtering equations}:\footnote{The conditional distribution of $\V{u}$ given $\V{z}$ is denoted by $\V{u} | \V{z}$, and $i:j$ means ``observations from time instants $i$ to $j$'', both extremes included.}
\begin{align}
    &\text{Prediction step for hidden variables:} \nonumber \\
    &\quad\Vg{\theta}_n | \V{y}_{1:n - 1} \sim N(\V{a}_n, \V{R}_n) \label{eq:kf-pred-hid} \\
    &\text{Prediction step for observed variables:} \nonumber \\
    &\quad\V{y}_n | \V{y}_{1:n - 1} \sim N(\V{f}_n, \V{Q}_n) \label{eq:kf-pred-obs} \\
    &\text{Update step (compare predictions to measurements):} \nonumber \\
    &\quad\Vg{\theta}_n | \V{y}_{1:n} \sim N(\V{k}_n, \V{C}_n), \label{eq:kf-compare-w-meas}
\end{align}
where\footnote{The superscript $^T$ after a vector or matrix denotes its transpose; the superscript $^{-1}$ after a matrix denotes its inverse.}
\begin{align}
    \V{a}_n &= \V{G}_n\V{k}_{n - 1} \label{eq:kf-update-1st} \\
    \V{R}_n &= \V{G}_n\V{C}_{n - 1}\V{G}_n^T + \V{W}_n \\
    \V{f}_n &= \V{F}_n\V{a}_n \\
    \V{Q}_n &= \V{F}_n\V{R}_n\V{F}_n^T + \V{V}_n \\
    \V{k}_n &= \V{a}_n + \left[\V{R}_n\V{F}_n^T\V{Q}_n^{-1}\right]\V{e}_n \\
    \V{e}_n &= \V{y}_n - \V{f}_n \\
    \V{C}_n &= \V{R}_n - \left[\V{R}_n\V{F}_n^T\V{Q}_n^{-1}\right]\V{F}_n\V{R}_n, \label{eq:kf-update-last}
\end{align}
assuming that the initial state is chosen according to a normal distribution, that is, $\Vg{\theta}_0 \sim N(\V{k}_0, \V{C}_0)$. For more details on the KF, see \cite{petris-dyn-models-r, grewal-andrews-kf}.

One of the appealing aspects of the KF is its ability to perform estimation and forecasting sequentially, as new data emerge. However, if observations $\V{y}_n$ for $n = 1, \dots, N$ are available beforehand, one is also able to retrospectively reconstruct the system's states, in order to analyze its behavior given all the observations. For this purpose, a backward-recursive algorithm can be employed to compute the conditional distributions of $\Vg{\theta}_n$ given $\V{y}_{1:N}$, for any $n < N$ \cite{petris-dyn-models-r, grewal-andrews-kf}. The main ingredient of this algorithm is the \textit{smoothing equation} \eqref{eq:kf-smooth}:
\begin{equation}
    \Vg{\theta}_n | \V{y}_{1:N} \sim N(\V{s}_n, \V{S}_n), \label{eq:kf-smooth}
\end{equation} 
where
\begin{align}
    \V{s}_n &= \V{k}_n + \V{C}_n\V{G}_{n + 1}^T\V{R}_{n + 1}^{-1}\left[\V{s}_{n + 1} - \V{a}_{n + 1}\right] \\
    \V{S}_n &= \V{C}_n - \V{C}_n\V{G}_{n + 1}^T\V{R}_{n + 1}^{-1} \times \nonumber \\
    & \quad\quad\quad\quad \left[\V{R}_{n + 1} - \V{S}_{n + 1}\right]\V{R}_{n + 1}^{-1}\V{G}_{n + 1}\V{C}_n,
\end{align}
assuming that $\Vg{\theta}_{n + 1} | \V{y}_{1:N} \sim N(\V{s}_{n + 1}, \V{S}_{n + 1})$. Notice that since the smoothing is performed backwards, it is necessarily to previously filter the set of observations to gain access to vectors $\V{k}_n$ and $\V{a}_n$, and matrices $\V{C}_n$ and $\V{R}_n$.

\section{Kalman Filter model for ensemble synchronization} \label{sec:kf-model}
Equations \eqref{eq:wing-2014-model-dynamic-ab-t}, \eqref{eq:wing-2014-model-dynamic-ab-T}, \eqref{eq:wing-2014-model-dynamic-ab-alpha}, and \eqref{eq:wing-2014-model-dynamic-ab-beta} can be written as a KF by considering proper choices for the observed and hidden variables, as well as the observation and state-transition matrices. The main goal of this section is to construct a sequence of matrices $\V{F}_n$ and $\V{G}_n$, as well as vectors $\V{y}_n$ and $\Vg{\theta}_n$ of observed and hidden variables respectively, such that Equations \eqref{eq:kf-obs} and \eqref{eq:kf-hidden} recover the model proposed in Equations \eqref{eq:wing-2014-model-dynamic-ab-t}, \eqref{eq:wing-2014-model-dynamic-ab-T}, \eqref{eq:wing-2014-model-dynamic-ab-alpha}, and \eqref{eq:wing-2014-model-dynamic-ab-beta}. Firstly, to simplify the formulation of the model, the observed variables are not the tone onset times for each player, but rather the \textit{inter-onset-intervals} (IOIs), denoted by $r_{i, k} = t_{i, n} - t_{i, n - 1}$, for $i = 1, \dots, K$. These values are assembled as in Equation \eqref{eq:kf-y}:
\begin{equation}
    \V{y}_n = \left[r_{1, n} ~ \dots ~ r_{K, n}\right] \in \RR^K \label{eq:kf-y}.
\end{equation}
The hidden variable $\Vg{\theta}_n$ can be written as in Equation \eqref{eq:kf-theta}:
\begin{equation}
    \Vg{\theta}_n = \left[\V{T}_n^T ~ \middle| ~ \V{r}_n^T ~ \middle| ~ \Vg{\alpha}_n^T ~ \middle| ~ \Vg{\beta}_n^T \right]^T \in \RR^{2K^2}, \label{eq:kf-theta}
\end{equation}
where
\begin{align}
    \V{T}_n &= \left[T_{1, n} ~ \dots ~ t_{K, n}\right]^T \in \RR^K \label{eq:kf-theta-T} \\
    \V{r}_n &= \left[r_{1, n} ~ \dots ~ r_{K, n}\right]^T \in \RR^K \label{eq:kf-theta-r} \\
    \Vg{\alpha}_n &= \left[\alpha_{ij, n} \text{ in the lexicographical order on } ij, \right. \nonumber \\ 
            &\quad ~~ \left.\text{for } 1 \leq i, j \leq K, i \neq j\right] \in \RR^{K(K - 1)} \label{eq:kf-theta-alpha} \\
    \Vg{\beta}_n &= \left[\beta_{ij, n} \text{ in the lexicographical order on } ij, \right. \nonumber \\ 
            &\quad ~~ \left.\text{for } 1 \leq i, j \leq K, i \neq j\right] \in \RR^{K(K - 1)}. \label{eq:kf-theta-beta}
\end{align}
The relation between $\Vg{\theta}_n$ and $\V{y}_n$ is described by the observation matrix in Equation \eqref{eq:kf-F}:\footnote{The identity matrix of dimensions $L \times L$ is denoted by $\V{I}_L$; the matrix of dimensions $L \times M$ filled with zeros is denoted by $\V{0}_{L \times M}$; a square null matrix of dimensions $L \times L$ is abbreviated by $\V{0}_L.$}
\begin{equation}
    \V{F}_n = \left[\V{0}_K ~ \middle| ~ \V{I}_K ~\middle| ~\V{0}_{K \times K(K - 1)} ~ \middle| ~ \V{0}_{K \times K(K - 1)} \right]. \label{eq:kf-F}
\end{equation}
Notice that matrices $\V{F}_n \in \RR^{K \times 2K^2}$ are constant through time. The evolution of the hidden variables in $\Vg{\theta}_n$ is modelled by a sequence of state-transition matrices $\V{G}_n \in \RR^{2K^2 \times 2K^2}$, described in Equation \eqref{eq:kf-G}:
\begin{equation}
    \begingroup
    \setlength\arraycolsep{1.3pt}
    \!\!\!
    \begin{bmatrix}
        \V{I}_K & \V{0}_K & \V{0}_{K \times K(K - 1)} & \V{G}_n^{T\beta} \\
        \V{I}_K & \V{0}_K & \V{G}_n^{r\alpha} & \V{G}_n^{r\beta} \\
        \V{0}_{K(K - 1) \times K} & \V{0}_{K(K - 1) \times K} & \V{I}_{K(K - 1)} & \V{0}_{K(K - 1)} \\
        \V{0}_{K(K - 1) \times K} & \V{0}_{K(K - 1) \times K} & \V{0}_{K(K - 1)} & \V{I}_{K(K - 1)}
    \end{bmatrix}\!\!, \label{eq:kf-G}
    \endgroup
\end{equation}
where matrices $\V{G}_n^{T\beta}$, $\V{G}_n^{r\alpha}$, and $\V{G}_n^{r\beta}$ (of dimensions $K \times K(K - 1)$ each) describe the interaction between variables in their respective superscripts. These three matrices are equal to the matrix in Equation \eqref{eq:kf-G-inter}:
\begin{equation}
    \begin{bmatrix}
        - \V{A}_{1:, n - 1}^T & \V{0}_{1 \times (K - 1)} & \cdots & \V{0}_{1 \times (K - 1)} \\
        \V{0}_{1 \times (K - 1)} & - \V{A}_{2:, n - 1}^T & \cdots & \V{0}_{1 \times (K - 1)} \\
        \vdots & \vdots & \ddots & \vdots \\
        \V{0}_{1 \times (K - 1)} & \V{0}_{1 \times (K - 1)} & \cdots & - \V{A}_{K:, n - 1}^T
    \end{bmatrix}, \label{eq:kf-G-inter}
\end{equation}
where each $\V{A}_{i:, n - 1} \in \RR^{K - 1}$ contain the asynchronies $A_{ij, n - 1}$ of player $i$ to all players $j$, for $j \neq i$, at time $n - 1$. Vector $\V{A}_{i:, n - 1}$ is made explicit in Equation \eqref{eq:kf-A-vecs}: 
\begin{equation}
    \!\!\left[A_{i1, n - 1} \dots A_{i(i - 1), n - 1} A_{i(i +1), n - 1} \dots A_{iK, n - 1}\right]^T\!\!. \label{eq:kf-A-vecs}
\end{equation}

A simple (but tedious) verification using Equations \eqref{eq:kf-obs} and \eqref{eq:kf-hidden} with these choices for $\V{F}_n$, $\V{G}_n$, $\V{y}_n$, and $\Vg{\theta}_n$ ensures that the model in Equations 
\eqref{eq:wing-2014-model-dynamic-ab-t}, \eqref{eq:wing-2014-model-dynamic-ab-T}, \eqref{eq:wing-2014-model-dynamic-ab-alpha}, and \eqref{eq:wing-2014-model-dynamic-ab-beta} is recovered. It is also established that when $K = 1$ the model in Equations \eqref{eq:vorberg-schulze-model} and \eqref{eq:period-correction} is recovered, with the improvement of dynamic $\alpha$ and $\beta$.

When compared to the bGLS algorithm \cite{jacoby-bgls-1, jacoby-bgls-2}, the state-of-the-art to estimate parameters in the scenario of sensorimotor synchronization, the KF model presents a great advantage, that is the possibility of performing online estimation as more data become
available: this feature can be important if one desires to implement real-time synchronization schemes. When the complete time-series of onset times/IOIs is available, one can apply the smoothing equation \eqref{eq:kf-smooth}, in order to dynamically estimate the parameters of interest throughout the performance, as well as estimate them by applying the filtering equations \eqref{eq:kf-pred-hid}, \eqref{eq:kf-pred-obs}, and \eqref{eq:kf-compare-w-meas}, for example, to simulate an online scenario.

Notice that the dimensions of $\V{F}_n$, $\V{G}_n$, and $\Vg{\theta}_n$ scale quadratically with the number of performers, which may render the model overly complicated or cause computational issues when computing the KF update/filtering equations.\footnote{Other computational issues on the Kalman Filter are largely discussed in \cite{petris-dyn-models-r}.} However, due to the sparsness of matrices $\V{F}_n$ and $\V{G}_n$, block-multiplication will highly reduce the number of operations when computing Equations \eqref{eq:kf-update-1st} to \eqref{eq:kf-update-last}, mitigating the latter issue. Regarding the complexity of the model, notice that in real large-scale scenarios (eg. a symphony orchestra) it is not realistic to assume that each musician synchronizes with every other, thus allowing for potential simplifications, like considering a group of instruments as a single unity and synchronizing with every other group. This procedure would diminish the value of $K$ from approximately 100 to less than 20. A useful topic  for future research would be to investigate the possibility of modeling the synchronization scheme between performers (or group of performers) in a graph, in order to decrease even more the number of relevant connections.

\begin{figure*}[ht]
    \centering
    \includegraphics[width=1\textwidth]{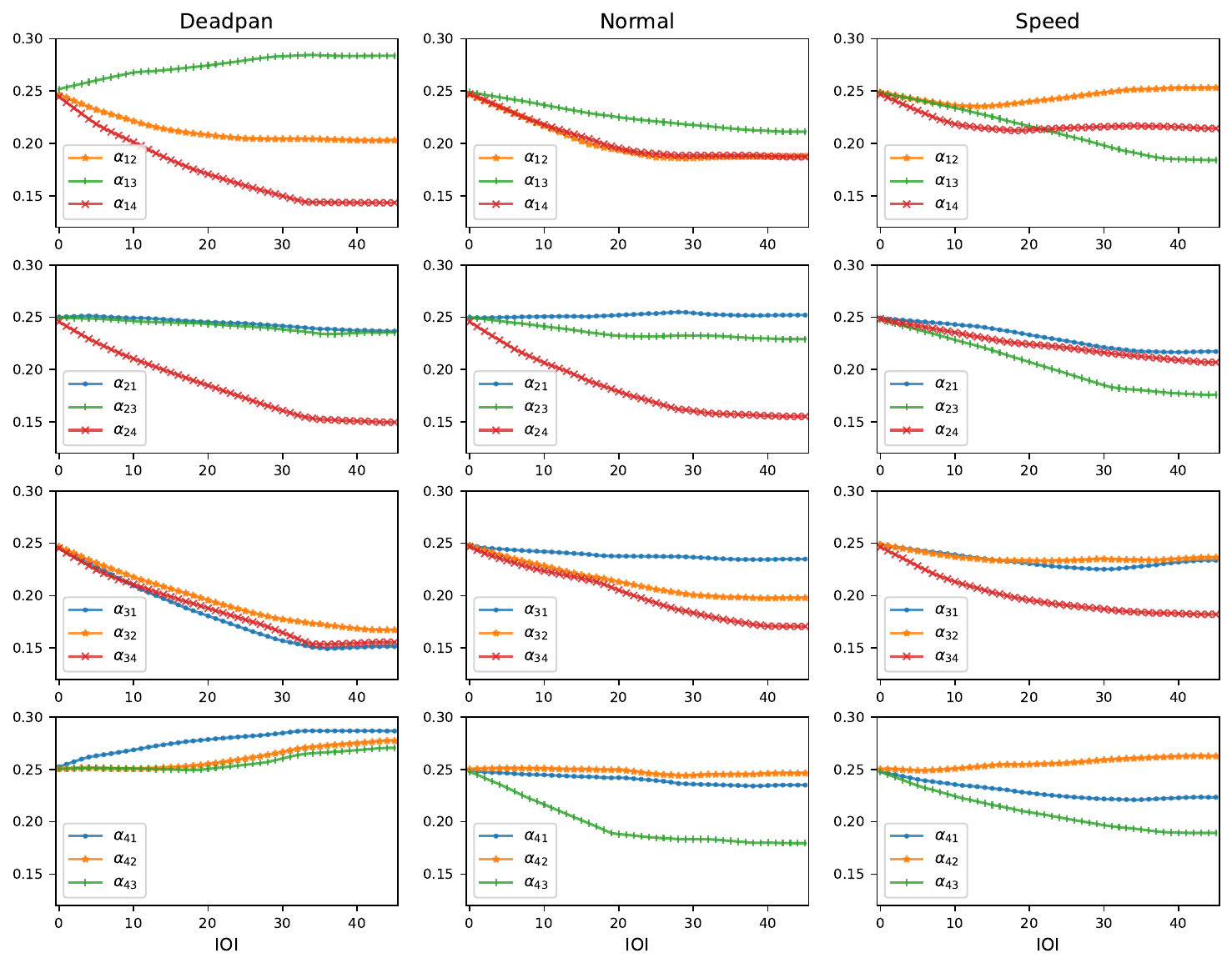}
    \caption{Smoothed time-series for the phase correction gain on three performance styles of an excerpt of the fourth movement of the string quartet Op. 74 no. 1, by Joseph Haydn. See Section \ref{sec:results} for discussion.}
    \label{fig:experiment}
\end{figure*}

Another issue that is important to point out is the design of the covariance matrices for the observation and process noises, $\V{V}_n$ and $\V{W}_n$ respectively. On a first view, it makes sense to consider $\V{V}_n$ as diagonal matrices, for simplicity, since the interaction between the performers is already ``captured''~by the correction gains in the hidden variables; however, it is not clear if $\V{W}_n$ should be a sequence of diagonal matrices, since it makes sense to consider at least correlations between both correction gains of the same performer. This work employs a particular choice for these covariance matrices, as will be further discussed in Section \ref{sec:results}. Further investigation on this question could involve coupling the Expectation-Maximization algorithm with the KF in order to estimate not only matrices $\V{V}_n$ and $\V{W}_n$ but also $\V{F}_n$ and $\V{G}_n$ \cite{mader-etal-em-kf}. However a disadvantage would be that these estimates would need to
be static through time, requiring a large amount of data, and being highly dependent of the piece of music being analyzed.

\section{Results}\label{sec:results}

To illustrate the effectiveness of the proposed model, a set of simulations was performed, using an excerpt of the dataset presented in \cite{virtuoso}, similar to the one used in \cite{wing-etal-2014}: the homophonic section from the fourth movement of the string quartet Op. 74 no. 1 by Joseph Haydn, from bars 13 to 24. In this excerpt the instruments play a sequence of 47 quarter notes in rhytmic unison, with the first violin breaking the pattern near the end with an adornment of four sixteenth notes, which are disregarded in this study. 

Three performance styles are considered: \textit{Normal} condition (concert-style performance); \textit{Speed} condition (including a spontaneous \textit{accelerando} and \textit{ritardando} initiated by a single musician -- the designated leader, that can be the first or second violin); and \textit{Deadpan} condition (performances with minimal expression in tempo and articulation). All the simulation were performed on a computer equipped with a 12th Generation Intel Core\texttrademark~i7 processor and 16GB of RAM, running Windows 11 Pro\texttrademark; the implementations were conducted in Python version 3.11.7.\footnote{Codes available at https://github.com/arme-project/ismir-2024.}

Regarding the parameters of the KF, the covariance matrix for the process noise, $\V{W}_n$, plays an important role, since it indicates how the variables in $\Vg{\theta}$ interact. Based on the interpretation of the hidden variables, a reasonable choice for all the $\V{W}_n$ is the block-diagonal matrix in Equation \eqref{eq:W-block-diag}:\footnote{Off-diagonal blocks are null matrices, that were omitted exceptionally here, to avoid a line-break in the number of the equation. Moreover, notation $\V{W}^{(T)}$ means to avoid confusion with the transpose matrix.}
\begin{equation}
    \begin{bmatrix}
        \V{W}^{(T)} & & & \\
        & \V{W}^r & & \\
        & & \V{W}^{\alpha} & \\
        & & & \V{W}^{\beta} 
    \end{bmatrix},
    \label{eq:W-block-diag}
\end{equation}
where $\V{W}^{(T)}$ and $\V{W}^r$ are given respectively by $\sigma_T^2 \V{I}_K$ and $\sigma_r^2 \V{I}_K$, being $\sigma_T^2$ the \textit{timekeeper variance} and $\sigma_r^2$ the \textit{motor variance}. Since it is known that the motor variance is way smaller than the timekeeper variance \cite{wing-etal-2014, jacoby-bgls-1, jacoby-bgls-2}, the conservatively high values $\sigma_T^2 = 500$ and $\sigma_r^2 = 25$ were considered. Both $\V{W}^{\alpha}$ and $\V{W}^{\beta}$ are also block-diagonal matrices, consisting of $K$ blocks, each of dimensions $(K - 1) \times (K - 1)$ and as in Equation \eqref{eq:W-ab}:
\begin{equation}
    \begin{bmatrix}
        v & c & \cdots & c \\
        c & v & \cdots & c \\
        \vdots & \vdots & \ddots & \vdots \\
        c & c & \cdots & v
    \end{bmatrix},
    \label{eq:W-ab}
\end{equation}
where $v$ represents the variance of each $\alpha_{ij}$ (or $\beta_{ij}$) and $c$ is the covariance between two distinct $\alpha_{ij}$ (or $\beta_{ij}$).

The rationale behind this construction for $\V{W}^{\alpha}$ and $\V{W}^{\beta}$ is that it makes sense to assume that for performer $i$ there is a correlation only between the $\alpha_{ij}$ (or $\beta_{ij}$) for $j \neq i$. This means that all the correction gains for performer $i$ interact among themselves, but not directly with the correction gains of other performers. Also, it is expected that the correlation between two distinct $\alpha_{ij}$ (or $\beta_{ij}$) is negative, since increasing correction towards a specific performer may cause a decrease of the synchronization towards the others. With this in mind, for matrix $\V{W}^{\alpha}$ the value of $v$ was chosen as $10^{-4}$; the value of $c$ was chosen such that the correlation between any two distinct $\alpha_{ij}$ is equal to $-0.1$.\footnote{This procedure will not always lead to a positive-definite matrix, for sufficiently high value of $K$ and depending on $c$ -- not the case here.}

In this preliminary set of experiments with the KF, the effect of the $\beta_{ij}$ was disregarded, by considering $\V{W}^{\beta}$ a null matrix. It is known that the effect of the phase correction is way more relevant than the effect of the period correction \cite{wing-etal-2014, jacoby-bgls-1, jacoby-bgls-2}, with the $\beta_{ij}$ coefficients being usually much smaller than the $\alpha_{ij}$. Also, preliminary experiments with artificial data also indicate that the dynamic values of the phase correction may render the period correction unnecessary. Since this is a point to be further investigated, it seemed safe to first experiment only with phase correction.

Matrices $\V{V}_n$ were chosen to be constant and equal to $10^{-5} \V{I}_K$: since the block $\V{W}^r$ in matrix $\V{W}_n$ already captures the motor variance, $\V{V}_n$ should be a sequence of null matrices, but a negligible diagonal term was added to avoid numerical errors. Finally, the initialization of vector $\Vg{\theta}$ was done by choosing its first $K$ components and the components from $K + 1$ to $2K$ to be equal to the first IOI of each of the $K$ instruments, all the $\alpha_{ij}$ were initially set to $0.25$, and all the $\beta_{ij}$ to zero. This initialization of $\alpha_{ij}$ is supported by \cite{wing-etal-2014}, where optimal correction values for ensembles of size $K$ were derived.

Figure \ref{fig:experiment} summarizes one experiment performed in the aforementioned scenario. Three repetitions of the Haydn quartet excerpt were analyzed, being one for each of the three performance conditions, having the second violin as the leader in the ``Speed''~case. Each performance consists of a sequence of 46 four-dimensional vectors containing the IOIs for each instrument. Since it is not the goal of this set of experiments to evaluate online performance of the proposed model, these three sequences were smoothed by the KF,\footnote{The computational time of each smoothing is less than 100ms.} according to Equation \ref{eq:kf-smooth}. Each panel of Figure \ref{fig:experiment} displays the evolution of the $\alpha_{ij}$, for $j \neq i$, organized as follows: each column contains a performance condition (made explicit at its top), and each row displays the evolution of $\alpha_{ij}$ for $j \neq i$ and a fixed value of $i$. The conditioning of each $\alpha_{ij}$ on $\V{y}_{1:N}$ is omitted, and the instruments are abbreviated by numbers, where 1, 2, 3, and 4 refers to the first violin, second violin, viola, and cello, respectively. On each panel of Figure \ref{fig:experiment} the values of $\alpha_{ij}$ promptly deviates from the optimal initialization of $0.25$ (but still varies around it), and their respective behavior are now discussed.

In ``Speed''~condition (third column in Figure \ref{fig:experiment}), on each panel the phase correction parameter toward the second violin ($\alpha_{i2}$, for $i = 1, 3, 4$) shows a small increase by the end of the performance, when the change in speed occurs, since the second violin is assigned as the leader to initiate this change in speed. Notice also that his/her phase correction parameters towards the other performers ($\alpha_{2j}$, for $j = 1, 3, 4$) decrease through time, specially near the last notes, reinforcing its leadership in this tempo change.

In the ``Normal''~condition (second column in Figure \ref{fig:experiment}) it is noticeable that the second violin, viola, and cello are systematically synchronizing mainly to the first violin, which plays the melody in this excerpt: notice the almost constant value for $\alpha_{i1}$, for $i = 2, 3, 4$. While the cello is synchronizing mainly with the first and second violin, it presents the weaker ``synchronization attractor'', as seen by the significant decrease in $\alpha_{i4}$ through time, for $i = 1, 2, 3$.

Finally, in the ``Deadpan''~condition (first column in Figure \ref{fig:experiment}) the first and second violin and the cello are synchronizing mainly to the viola (steady increase of $\alpha_{i3}$ for $i = 1, 2, 4$, and decrease in $\alpha_{3j}$ for $j = 1, 2, 4$), which may be the cause of the cello synchronizing systematically with all the three other instruments.

This experiment indicates that the proposed model is capable of capturing local fluctuations in tempo, reinforces the role of the phase correction gain in interpreting synchronization mechanisms in musical ensembles, and assess qualitatively the validity of a time-varying model to the problem of ensemble synchronization. As a next  step in this new direction for the field, the proposed model will be broadly tested and systematically compared with other models. Some issues to be addressed in future work are: perform experiments with other data contained on \cite{virtuoso}; compare filtering and smoothing procedures, as well as investigate if the filtered estimates make sense from a music cognition perspective; implement tools from the theory of dynamic linear models to automatically estimate the covariance matrices $\V{V}_n$ and $\V{W}_n$ \cite{petris-dyn-models-r}; perform a systematic comparison with the bGLS and ADAM algorithms.


\section{Conclusion}\label{sec:conc-fut-works}
This paper presented a novel model, based on the Kalman Filter, for analysing asynchrony correction in music ensemble performances. The proposed model is founded on well-established models in the literature, and has the advantage of considering dynamic phase and period correction gains. A set of experiments (using only phase correction) on a homophonic section of a string quartet by J. Haydn was conducted, illustrating the capabilities of the model in explaining synchronization schemes within musical ensembles.

\section{Acknowledgments}
The ARME Project (Augmented Reality Music Ensemble -- https://arme-project.co.uk/) is funded by the EPSRC grant with reference EP/V034987/1. We would like to thank the University of Birmingham for the scholarship awarded to the first author through the Brazil Visiting Fellows Scheme. We also would like to thank the reviewers and meta-reviewers for the useful insights and suggestions to improve this paper.

\bibliography{ISMIRtemplate}

\begin{thebibliography}{10}
\providecommand{\url}[1]{#1}
\csname url@samestyle\endcsname
\providecommand{\newblock}{\relax}
\providecommand{\bibinfo}[2]{#2}
\providecommand{\BIBentrySTDinterwordspacing}{\spaceskip=0pt\relax}
\providecommand{\BIBentryALTinterwordstretchfactor}{4}
\providecommand{\BIBentryALTinterwordspacing}{\spaceskip=\fontdimen2\font plus
\BIBentryALTinterwordstretchfactor\fontdimen3\font minus \fontdimen4\font\relax}
\providecommand{\BIBforeignlanguage}[2]{{%
\expandafter\ifx\csname l@#1\endcsname\relax
\typeout{** WARNING: IEEEtran.bst: No hyphenation pattern has been}%
\typeout{** loaded for the language `#1'. Using the pattern for}%
\typeout{** the default language instead.}%
\else
\language=\csname l@#1\endcsname
\fi
#2}}
\providecommand{\BIBdecl}{\relax}
\BIBdecl

\bibitem{patel-etal}
A.~D. Patel, J.~R. Iversen, M.~R. Bregman, and I.~Schulz, ``Experimental evidence for synchronization to a musical beat in a nonhuman animal,'' \emph{Current Biology}, vol.~19, no.~10, pp. 827--830, 2009.

\bibitem{wing-woodburn}
A.~M. Wing and C.~Woodburn, ``The coordination and consistency of rowers in a racing eight,'' \emph{Journal of Sports Sciences}, vol.~13, no.~3, pp. 187--197, 1995.

\bibitem{goodman}
E.~Goodman, ``Ensemble performance,'' in \emph{Musical performance: a guide to understanding}, J.~Rink, Ed.\hskip 1em plus 0.5em minus 0.4em\relax Cambridge, UK: Cambridge University Press, 2002, pp. 153--167.

\bibitem{palmer}
C.~Palmer, ``Music performance,'' \emph{Annual Review of Psychology}, vol.~48, pp. 115--138, 1997.

\bibitem{davidson-good}
J.~W. Davidson and J.~M.~M. Good, ``Social and musical co-ordination between members of a string quartet: An exploratory study,'' \emph{Psychology of Music}, vol.~30, no.~2, pp. 186--201, 2002.

\bibitem{murninghan-conlon}
J.~K. Murnighan and D.~E. Conlon, ``The dynamics of intense work groups: A study of british string quartets,'' \emph{Administrative Science Quarterly}, vol.~36, no.~2, pp. 165--186, 1991.

\bibitem{gibbon-et-al}
J.~Gibbon, C.~Malapani, C.~L. Dale, and C.~R. Gallistel, ``Toward a neurobiology of temporal cognition: advances and challenges,'' \emph{Current Opinion in Neurobiology}, vol.~7, no.~2, pp. 170--184, 1997.

\bibitem{wing-etal-2014}
A.~M. Wing, S.~Endo, A.~Bradbury, and D.~Vorberg, ``Optimal feedback correction in string quartet synchronization,'' \emph{Journal of the Royal Society Interface}, vol.~11, no. 20131125, 2014.

\bibitem{vorberg-schulze}
D.~Vorberg and H.~H. Schulze, ``Linear phase correction in synchronization: predictions, parameter estimation, and simulations,'' \emph{Journal of Mathematical Psychology}, vol.~46, no.~1, pp. 56--87, 2002.

\bibitem{vorberg-wing}
D.~Vorberg and A.~M. Wing, ``Modeling variability and dependence in timing,'' in \emph{Handbook of perception and action, vol. 2}, H.~Heuer and S.~Keele, Eds.\hskip 1em plus 0.5em minus 0.4em\relax New York, USA: Academid Press, 1996, pp. 181--262.

\bibitem{mates}
J.~Mates, ``A model of synchronization of motor acts to a stimulus sequence,'' \emph{Biological Cybernetics}, vol.~71, p. 186, 1994.

\bibitem{repp-adap-tempo-changes}
B.~H. Repp, ``Processes underlying adaptation to tempo changes in sensorimotor synchronization,'' \emph{Human Movement Science}, vol.~20, no.~3, pp. 277--312, 2001.

\bibitem{repp-keller-adap-tempo-changes}
B.~H. Repp and P.~E. Keller, ``Adaptation to tempo changes in sensorimotor synchronization: effects of intention, attention, and awareness,'' \emph{Quarterly Journal of Experimental Psychology}, vol.~57, no.~3, pp. 499--521, 2004.

\bibitem{jacoby-bgls-1}
N.~Jacoby, N.~Tishby, B.~H. Repp, M.~Ahissar, and P.~E. Keller, ``Parameter estimation of linear sensorimotor synchronization models: Phase correction, period correction, and ensemble synchronization,'' \emph{Timing \& Time Perception}, vol.~3, no. 1--2, pp. 52--87, 2015.

\bibitem{jacoby-bgls-2}
N.~Jacoby, P.~E. Keller, B.~H. Repp, M.~Ahissar, and N.~Tishby, ``Lower bound on the accuracy of parameter estimation methods for linear sensorimotor synchronization models,'' \emph{Timing \& Time Perception}, vol.~3, no. 1--2, pp. 32--51, 2015.

\bibitem{adam-2013}
M.~C. van~der Steen and P.~E. Keller, ``{The ADaptation and Anticipation Model (ADAM) of sensorimotor synchronization},'' \emph{Frontiers in Human Neuroscience}, vol.~7, 2013.

\bibitem{adam-2019}
B.~Harry and P.~E. Keller, ``{Tutorial and simulations with ADAM: an adaptation and anticipation model of sensorimotor synchronization},'' \emph{Biological Cybernetics}, vol. 113, pp. 397--421, 2019.

\bibitem{petris-dyn-models-r}
G.~Petris, S.~Petrone, and P.~Campagnoli, \emph{Dynamic Linear Models with R}.\hskip 1em plus 0.5em minus 0.4em\relax Berlin/Heidelberg, Germany: Springer, 2009.

\bibitem{grewal-andrews-kf}
M.~S. Grewal and A.~P. Andrews, \emph{Kalman Filtering: Theory and Practice with MATLAB}.\hskip 1em plus 0.5em minus 0.4em\relax New Jersey, USA: Wiley-IEEE Press, 2014.

\bibitem{mader-etal-em-kf}
W.~Mader, Y.~Linke, M.~Mader, L.~Sommerlade, J.~Timmer, and B.~Schelter, ``A numerically efficient implementation of the expectation maximization algorithm for state space models,'' \emph{Applied Mathematics and Computation}, vol. 241, pp. 222--232, 2014.

\bibitem{virtuoso}
M.~Tomczak, M.~S. Li, and M.~D. Luca, ``Virtuoso strings: A dataset of string ensemble recordings and onset annotations for timing analysis,'' in \emph{Extended Abstracts for the Late-Breaking Demo Session of the 24th Int. Society for Music Information Retrieval Conf.}, Milan, Italy, 2023.

\end{thebibliography}

\end{document}